\documentclass[11pt]{elsart}
\usepackage[dvips]{graphicx}
\usepackage{amsmath}
\usepackage{array}
\usepackage{amssymb}
\usepackage{amsmath}
\usepackage{hhline}
\usepackage{longtable}

\begin{document}
\begin{frontmatter}

\title{Correlations in random Apollonian network}
\author{Zhongzhi Zhang}\quad
\ead{zhangzz@fudan.edu.cn}
\author{Shuigeng Zhou\corauthref{zz}}\corauth[zz]
{Corresponding author.}
\ead{sgzhou@fudan.edu.cn}
\address{
Department of Computer Science and Engineering, \\ Fudan
University, Shanghai 200433, China}%
\address{
Shanghai Key Lab of Intelligent Information Processing,\\ Fudan
University, Shanghai 200433, China}%

\begin{abstract}
In this paper, by both simulations and theoretical predictions we
study two and three node (or degree) correlations in random
Apollonian network (RAN), which have small-world and scale-free
topologies. Using the rate equation approach under the assumption of
continuous degree, we first give the analytical solution for two
node correlations, expressed by average nearest-neighbor degree
(ANND). Then, we revisit the degree distribution of RAN using rate
equation method and get the exact connection distribution, based on
which we derive a more accurate result for mean clustering
coefficient as an average quantity of three degree correlations than
the one previously reported. Analytical results reveal that ANND has
no correlations with respect to degree, while clustering coefficient
is dependent on degree, showing a power-law behavior as $C(k)\sim
k^{-1}$. The obtained expressions are successfully contrasted with
extensive numerical simulations.
\begin{keyword}
Complex networks\sep Correlations\sep Scale-free networks\sep
Small-world networks\sep  Networks %\PACS 89.75.Hc,
%89.75.Fb,05.10.-a, 87.23.Kg
\end{keyword}
\end{abstract}
%\pacs{89.20.Hh, 89.75.Hc, 89.75.Da}
%89.20.Hh World Wide Web, Internet
%89.75.Da Systems obeying scaling laws
%89.75.Fb Structures and organization in complex systems
%89.75.-k Complex systems
%89.75.Hc Networks and genealogical trees

\date{}
\end{frontmatter}

%%%%%%%%%%%%%%%%%%%%%%%%%%%%%%%%%%%%%%%%%%%%%%%%%%%%%%%%%%%%%%%%%
%%%%%%%%%%%%%%%%%%%%%%%%%%%%%%%%%%%%%%%%%%%%%%%%%%%%%%%%%%%%%%%%%
%\vskip -0.5cm\color{Blue}
%\vbox to 0pt{\kern -14cm {
%\noindent \small \copyright 2005
%{\em Elsevier Science B.V. All rights reserved}\\
%{\em Physica A}, submitted.}
%\vss}\color{Black}

%%%%%%%%%%%%%%%%%%%%%%%%%%%%%%%%%%%%%%%%%%%%%%%%%%%%%%%%%%%%%%%%%%%%
\section{Introduction}
The last few years have witnessed the birth of a new movement of
interest and research in the study of complex networks as an
interdisciplinary subject
\cite{AlBa02,DoMe03,SaVe04,NeBaWa06,BoLaMoChHw06}. This flurry of
activity, triggered by two ground-breaking papers by Watts and
Strogatz on small-world networks \cite{WaSt98} and Barab\'asi and
Albert on scale-free networks \cite{BaAl99}, has been certainly
induced by the increased computing powers and by the possibility to
study the properties of a plenty of large databases of real-life
networks, such as Internet, World Wide Web, metabolic networks,
protein networks in the cell, co-author networks and citation
networks. These networks have been identified and analyzed in the
literature, the emphasis being mostly on the three basic topological
characteristics, such as power-law degree distribution, small
average path length (APL) and high clustering coefficient
\cite{AlBa02,DoMe03,SaVe04,NeBaWa06,BoLaMoChHw06}.

However, the above mentioned three properties do not provide
sufficient characterizations of the real-world systems. In fact, it
has been observed that real networks exhibit ubiquitous degree
correlations among their
nodes~\cite{MsSn02,PaVaVe01,VapaVe02,Newman02,Newman03c}. This
translates in the observation that the degrees of nearest neighbor
nodes are not statistically independent but mutually correlated in
most real-life networks. Correlations play an important role in the
characterization of  network topology, and have led to a first
classification of complex networks~\cite{Newman02}. They are thus a
very relevant issue, especially in view of the important
consequences that they can have on dynamical processes taking place
on networks~\cite{BoPa02,MoVa03,VaMo03,EcGaMoVa05}.

Recently, based on the well-known Apollonian packing, deterministic
Apollonian network (DAN)~\cite{AnHeAnSi05} and random Apollonian
network (RAN)~\cite{ZhYaWa05}, called jointly Apollonian networks
(ANs) were proposed, which have been generalized to high
dimension~\cite{DoMa05,ZhCoFeRo05,ZhRoCo05,ZhRoZh06}. ANs are
simultaneously scale-free, small-world, Euclidean (i.e. it can be
embedded in an Euclidean lattice~\cite{AnHeAnSi05}), space filling,
and with matching graphs (that is to say, in the infinite network
size limit, the edges do not only completely cover the space like a
Peano curve, but also never cross each other). They may provide
valuable insight into the real-life networks, e.g. the maximal
planarity~\cite{ZhYaWa05} of ANs is of particular practicability for
the layout of printed circuits. Moreover, ANs describe force chains
in polydisperse granular packings and could also be applied to the
geometry of fully fragmented porous media, hierarchical road
systems, and area-covering electrical supply
networks~\cite{AnHeAnSi05}. Particularly, DAN can be helpful to
understand the energy landscape networks~\cite{DoMa05,Do02,DoMa05b}.
Very recently, ANs have attracted increasing interest from the
scientific
community~\cite{AnHe05,LiGaHe04,AnMi05,HaMa06a,HaMa06b,HuXuWuWa06},
some interesting dynamical processes, such as percolation, epidemic
spreading, synchronization and random walks taking place on RAN have
been investigated.

In this paper we investigate the correlation properties of nodes'
connectivity of random Apollonian network
(RAN)~\cite{ZhYaWa05,ZhRoCo05,ZhRoZh06}. Combining a rate
equation~\cite{szAlKe03} in the continuous degree approximation and
the boundary condition~\cite{BoPa05} of this rate equation, we work
out analytically the two node correlations in RAN, measured by the
average nearest-neighbor degree (ANND) of nodes with degree $k$.
Then, using the genuine discrete degree distribution, we obtain the
exact analytical result of mean clustering coefficient, which is the
average value of three node correlations. Both the analytical
expressions are in very good agreement with numerical simulations.

\section{Random Apollonian network}

The random Apollonian network (RAN)~\cite{ZhYaWa05} is the
stochastic version of the deterministic Apollonian network
(DAN)~\cite{AnHeAnSi05}. The construction of DAN begins with ($t=0$)
a triangle and at each time step a new node is connected to the
nodes of every existing triangles, omitting those triangles that had
already been updated in previous steps. In the random version the
triangles to be updated are selected at random, one at a time. We
call the triangles to be updated \emph{active triangles}.

Here we focus on the random Apollonian network (RAN). Some
properties of the RAN have been investigated
\cite{ZhYaWa05,ZhRoCo05,ZhRoZh06}. The average degree of all its
nodes equals 6. It has a power-law degree distribution $P(k)\sim
k^{-3}$, and since each new node induces the addition of three
triangles, one expects that the RAN has a finite clustering
coefficient. It has been proved that the increasing tendency of
average path length of RAN is a little slower than the logarithm of
the number of nodes in RAN.

For the sake of the following investigation on correlations in RAN,
we first revisit the degree distribution of RAN. Since the network
size is incremented by one with each time step, we use the step
value $t$ to represent a node created at this step. Note that at
time $t$ there are $2t+1$ active triangles in RAN, and the number of
active triangles containing a node with degree $k$ is also equal to
$k$. Let $N_k(t)$ denote the average number of nodes with degree $k$
at time $t$. When a new node enters the network, $N_k(t)$ changes as
\cite{KaReLe00}
\begin{equation}\label{sf rate}
\frac{dN_k}{dt}=\frac{(k-1) N_{k-1}(t)-k N_k(t)}{2t+1}+\delta_{k,3}.
\end{equation}
In the asymptotic limit $N_k(t)=tP(k)$, where $P(k)$ is the degree
distribution. Eq. (\ref{sf rate}) leads to the following recursive
equation
\begin{equation}\label{recursive}
P(k)=\left\{\begin{array}{lcl}
\frac{k-1}{k+2}P(k-1) & \mbox{for} & k\geq 3+1\\
\frac{2}{5} & \mbox{for} & k=3\,,\\
\end{array}\right.
\end{equation}
giving
\begin{equation}\label{degree}
P(k)=\frac{24}{k(k+1)(k+2)}.
\end{equation}
In the limit of large $k$, $P(k)\sim k^{-3}$
\cite{ZhYaWa05,ZhRoCo05}, which has the same degree exponent as the
BA model \cite{BaAl99} and the hierarchical lattice
\cite{BeOs79,HiBe06}. This obtained degree distribution is a more
accurate result than the previous one \cite{ZhYaWa05,ZhRoCo05}.

Notice that the growing precess of RAN actually contains the
preferential attachment mechanism, which arises in it not because of
some special rule including a function of degree as in Ref.
\cite{BaAl99} but naturally. Indeed, the probability that a new node
created at time $t$ will be connected to an existing node $i$ is
clearly proportional to the number of active triangles containing
$i$, i.e. to its degree $k_i(t)$. Thus a node $i$ is selected with
the usual preferential attachment probability
$\Pi_{i}[k_i(t)]=k_i(t)/(2t+1) \sim k_i(t)/2t$ (for large $t$).
Consequently, $k_i$ satisfies the dynamical equation \cite{AlBa02}
\begin{equation}
\frac{\partial k_i(t)}{\partial t} = \frac{k_i(t)}{2t}.
\end{equation}
Considering the initial condition $k_i(i)=3$, we have
\cite{ZhRoCo05}
\begin{equation}\label{connect}
k_i(t)=3\left(\frac {t}{i}\right)^{1/2}.
\end{equation}
Equation (\ref{connect}) shows that the degree of all nodes evolves
the same way, following a power law as in the well-known BA
networks~\cite{BaAl99}.

\section{Correlations in Random Apollonian network}

Having obtained the exact degree distribution of the random
Apollonian networks (RAN), we now study the two and three node
correlations in RAN, which are merely two specific types of
correlations. We ignore the long-range and multinode correlations,
since the empirical data and theoretic research on such correlations
is also absent.

\subsection{Two node correlations}
Two node correlations in a network can be conveniently measured by
means of the quantity, called \emph{average nearest-neighbor degree}
(ANND), which is a function of node degree, and is more convenient
and practical in characterizing degree-degree correlations.  The
ANND is defined by \cite{PaVaVe01}
\begin{equation}
  k_{nn}(k) = \sum_{k'} k' P(k'|k).
  \label{knn1}
\end{equation}
If there are no two degree correlations, $k_{nn}(k)$ is independent
of $k$. When $k_{\rm nn}(k)$ increases (or decreases) with $k$, the
network is  is said to be assortative (or
disassortative)~\cite{Newman02,Newman03c}.

We can analytically compute the function value of $k_{nn}(k)$ for
the RAN. Let $R_i(t)$ denote the sum of the degrees of the neighbors
of node $i$, evaluated at time $t$. It is represented as
\begin{equation}
  R_i(t) =  \sum_{j \in V(i)} k_j(t),
  \label{Ri}
\end{equation}
where $ V(i)$ corresponds to the set of neighbors of node $i$. The
average degree of nearest neighbors of node $i$ at time $t$,
$k_{nn}(i,t)$, is then given by $k_{nn}(i,t) = R_i(t) / k_i(t)$.
During the growth of the RAN, $R_i(t)$ can only increase by the
addition of a new node connected either directly to $i$, or to one
of the neighbors of $i$. In the first case $R_i(t)$ increases by 3
(the degree of the newly created node), while in the second case it
increases by one unit. Therefore, in the continuous $k$
approximation \cite{AlBa02,DoMe03,SaVe04,NeBaWa06,BoLaMoChHw06}, we
can write down the following rate equation~\cite{szAlKe03,BoPa05}:
\begin{eqnarray}\label{dRi}
  \frac{dR_i(t)}{dt} = 3 \Pi_{i} [k_i(t)] + \sum_{j \in (V)(i)}
  \Pi_{j} [k_j(t)]  = \frac{3 k_i(t)}{2t} + \frac{R_i(t)}{2t} \ .
\end{eqnarray}
The general solution of Eq. (\ref{dRi}) is
\begin{equation}
  R_i(t)= \Phi_{0}(i) t^\frac{1}{2} +
  \frac{9}{2} \left( \frac{t}{i} \right)^{1/2} \ln t,
  \label{eq:12}
\end{equation}
where $\Phi_{0}(i)$ is determined by the boundary condition
$R_i(i)$. To obtain the boundary condition $R_i(i)$, we observe that
at time $i$, the new node $i$ is connected to an existing node $j$
of degree $k_j(i)$ with probability $\Pi_{j}[k_j(i)]$, and that the
degree of this node increase by one unit in the process. Thus,
\begin{equation}
  R_i (i) = \sum_{j=1}^{i}\Pi_{j} [k_j(i)] [k_j(i)+1].
  \label{eq:2}
\end{equation}
Inserting $\Pi_{j} [k_j(i)]=\frac{k_j(i)}{2i}$ and
$k_j(i)=3\left(\frac {i}{j}\right)^{1/2}$ into $R_i(i)$ leads to
\begin{equation}
  R_i (i) =  3 + \frac{9}{2} \sum_{j=1}^{i} \frac{1}{j}\simeq 3+\frac{9}{2}\ln i.
  \label{eq:11}
\end{equation}
So, in the large $i$ limit, $R_i (i)$ is dominated by the second
term, yielding
\begin{equation}
  R_i (i) =  \frac{9}{2}\ln i.
  \label{qq}
\end{equation}
From here, we have
\begin{equation}
  R_i(t) \simeq \frac{9}{2} \left(\frac {t}{i}\right)^{1/2} \ln t,
\end{equation}
and finally
\begin{equation}
  k_{nn}(k,t) \simeq  \frac{3}{2} \ln t.
\label{eq:knnBA}
\end{equation}
So, two node correlations do not depend on the degree. The ANND
grows with the network size $N=t$ as $\ln t$, in the same way as in
the BA model \cite{VapaVe02}.
%%%%%%%%%%%%%%%%%%%%%%%%%%%%%%%%%%%%%%%%%%%%%%%%%%%%%%%%%%
% Figure  1
%%%%%%%%%%%%%%%%%%%%%%%%%%%%%%%%%%%%%%%%%%%%%%%%%%%%%%%%%%
\begin{figure}
\begin{center}
      \begin{tabular}{cc}
        \includegraphics[width=6cm]{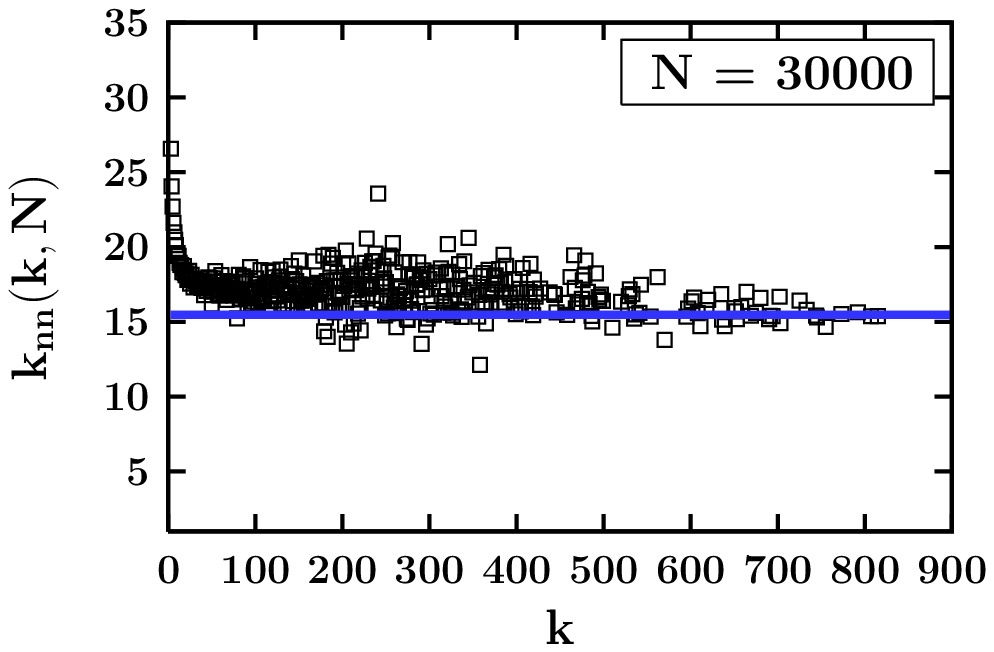}&\includegraphics[width=6cm]{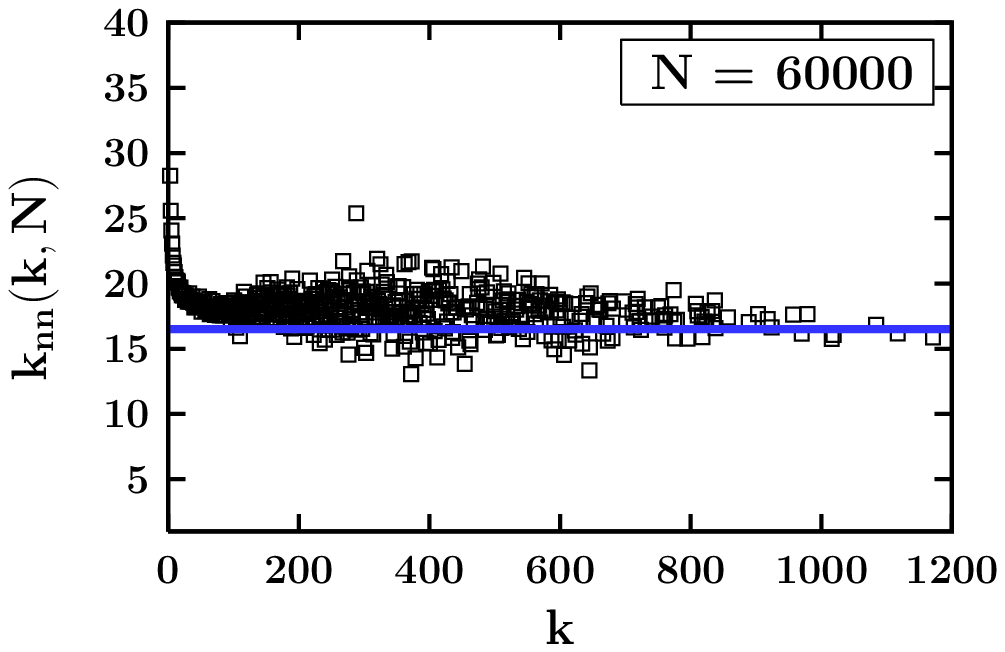}\\
        \includegraphics[width=6cm]{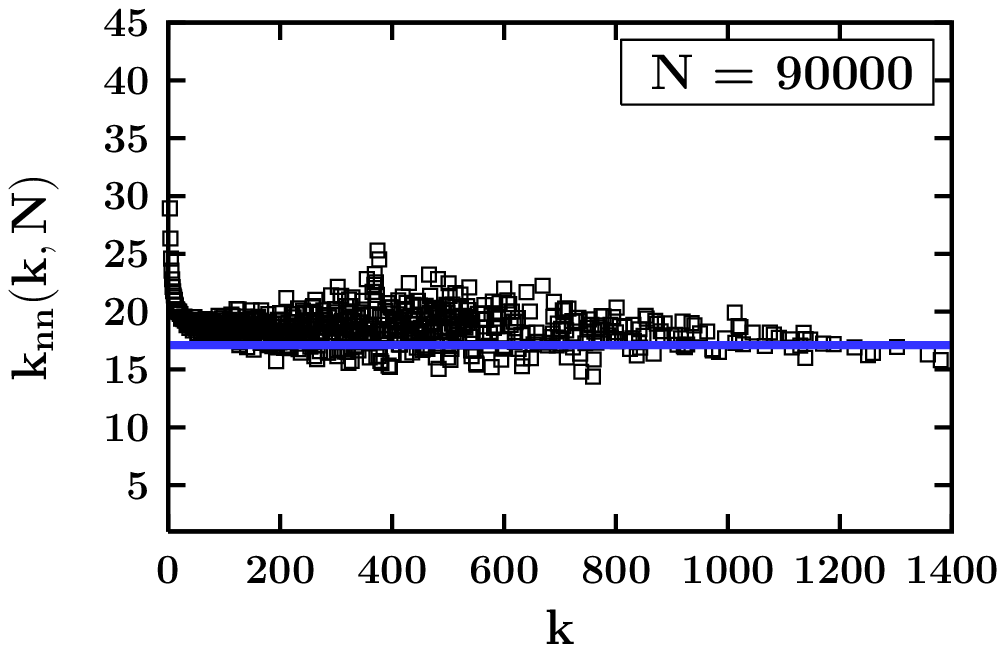}&\includegraphics[width=6cm]{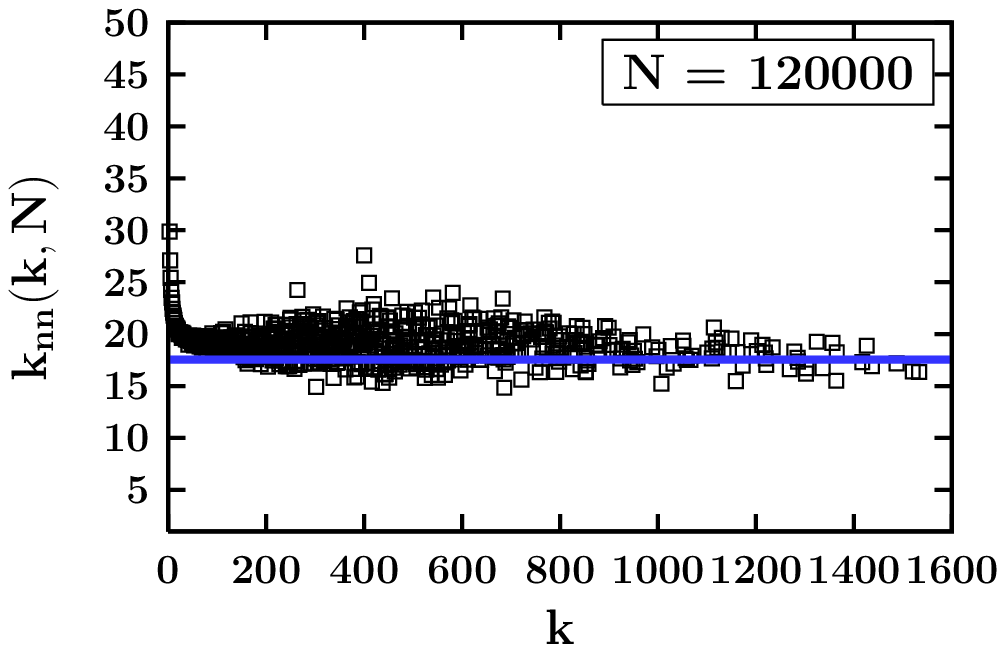}
    \end{tabular}
\caption{Plot
of average nearest-neighbor degree of the nodes with degree $k$. % at
%various network sizes $N$. The circles, squares, stars, and triangles
 The squares denote the simulation results for various network
 sizes, while the solid lines are the theoretical result provided by
Eq.~(\ref{eq:knnBA}).} \label{Fig1}
\end{center}
\end{figure}
%%%%%%%%%%%%%%%%%%%%%%%%%%%%%%%%%%%%%%%%%%%%%%%%%%%%%%%%%%

In order to confirm the validity of the obtained analytical
prediction of ANND, we performed extensive numerical simulations of
the RAN (see Fig.~\ref{Fig1}).
%Simulations were performed for different network sizes
%ranging from $N=10^3$ to $N=10^6$.
To reduce the effect of fluctuation on simulation results, %for every
%system with a certain size
the simulation results are average over
fifty network realizations. %Figure~\ref{Fig1} shows the average
%nearest neighbor degree with different numerical conditions.
From Fig.~\ref{Fig1} we observe that for large $k$ the ANND of
numerical and analytical results are in agreement with each other,
while the simulated results of ANND of small $k$ have a very weak
dependence on $k$, which is similar to the phenomena observed in the
BA model \cite{VapaVe02}. This $k$ dependence, for small degree,
cannot be detected by rate equation approach, since it has been
formulated in the continuous degree $k$ approximation.

%%%%%%%%%%%%%%%%%%%%%%%%%%%%%%%%%%%%%%%%%%%%%%%%%%%%%%%%%%
% Figure  2
%%%%%%%%%%%%%%%%%%%%%%%%%%%%%%%%%%%%%%%%%%%%%%%%%%%%%%%%%%
\begin{figure}
\begin{center}
  \includegraphics[width=9cm]{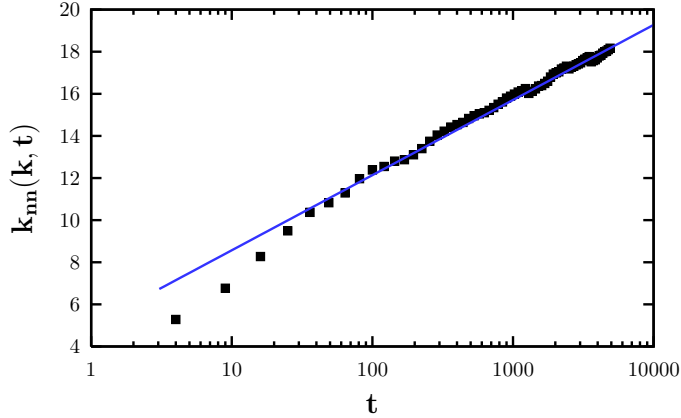}
  \caption{Semilogarithmic graph of time evolution for the ANND of the node added to the system at time
  2. The straight line follows $k_{nn}(k,t) = \frac{3}{2} \ln t$ as predicted by
Eq.~(\ref{eq:knnBA}). All data are from the average of 50
independent runs.} \label{Fig2}
\end{center}
\end{figure}
%%%%%%%%%%%%%%%%%%%%%%%%%%%%%%%%%%%%%%%%%%%%%%%%%%%%%%%%%%

From Eq. (\ref{eq:knnBA}), we can also easily know that the
evolution of ANND of all nodes is the same, showing a weak
logarithmic divergence with evolution time (network size). To
confirm this analytical prediction we also perform numerical
simulations. In Fig. \ref{Fig2} we report the numerical results,
which are in good agreement with this prediction provided by Eq.
(\ref{eq:knnBA}).

\subsection{Three node correlations}
Three node correlations can be measured by means of the conditional
probability $P(k', k''|k)$ that a node of degree $k$ is
simultaneously connected to nodes with degree $k'$ and $k''$. The
direct evaluation of the $P(k', k''|k)$ is generally difficult. To
overcome this problem, another interesting alternative quantity,
i.e. clustering coefficient, is frequently used. By definition, the
clustering coefficient~\cite{WaSt98} of nodes with degree $k$,
$C(k)$, is defined as the probability that two neighbors of a node
with degree $k$ are also neighbors themselves. The average
clustering coefficient of the whole network is then given as
\begin{equation}\label{ACC1}
  C = \sum_k P(k)C(k).
\end{equation}
Now we analytically estimate the average clustering coefficient $C$
of RAN by means of the clustering spectrum $C(k)$. In the RAN, for a
node of degree $k$, the exact value of its clustering coefficient
is~\cite{ZhYaWa05,ZhRoCo05,ZhRoZh06}
\begin{equation}\label{Ck}
C(k)=  \frac{4k-6}{k(k-1)},
\end{equation}
which depends on degree $k$. For large $k$, the degree-dependent
clustering $C(k)$ is inversely proportional to degree $k$, the same
behavior has been analytically found in some growing network models
such as deterministic (random) pseudofractal scale-free networks
and their variants
\cite{DoGoMe02,CoFeRa04,ZhRoZh06c,DoMeSa01,OzHuOt04,ZhRoGo05,ZhRoCo05a},
as well as some real systems \cite{RaBa03}.

\begin{figure}
\begin{center}
\includegraphics[width=9cm]{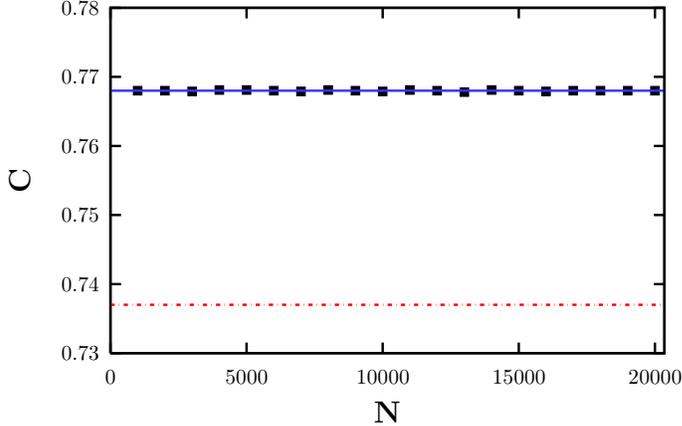}
\caption{Average clustering coefficient $C$ of RAN vs the network
size $N$. The solid line shows the exact result $\frac{4}{3}-
\left(119-12\pi^{2}\right)$, while the dashed line depicts the
previous analytic value $\frac{46}{3}-36\ln\frac{3}{2}$. The squares
denote the results of simulations.  Each data is obtained by fifty
independent network realizations.} \label{clu}
\end{center}
\end{figure}

Using Eq. (\ref{ACC1}), the average clustering coefficient $C$ of
RAN can be easily obtained as the mean value of $C(k)$ with respect
to the degree distribution $P(k)$ expressed by Eq. (\ref{degree}).
The result is
\begin{eqnarray}\label{ACC2}
 C &=&\sum_{k=3}^{\infty} P(k)C(k)=\sum_{k=3}^{\infty} \frac{24}{k(k+1)(k+2)} \frac{4k-6}{k(k-1)} \nonumber \\
 &=&\sum_{k=3}^{\infty} \frac{24}{k(k+1)(k+2)} \left(\frac{4}{k-1}-\frac{6}{k(k-1)}\right)= C_{1}- C_{2},
\end{eqnarray}
where $C_{1}=\sum_{k=3}^{\infty}
\frac{24}{k(k+1)(k+2)}\frac{4}{k-1}$ and $C_{2}=\sum_{k=3}^{\infty}
\frac{24}{k(k+1)(k+2)}\frac{6}{k(k-1)}$. We now compute in detail
$C_{1}$ and $C_{2}$, respectively. First, we can decompose $C_{1}$
into the sum of four terms as
\begin{eqnarray}\label{AC1}
 C_{1}
  =\sum_{k=3}^{\infty}
  \left(\frac{16}{k-1}-\frac{48}{k}+\frac{48}{k+1}-\frac{16}{k+2}\right)=\frac{4}{3}.
\end{eqnarray}
Analogously to Eq. (\ref{AC1}), we get
\begin{eqnarray}\label{AC2}
 C_{2}
  =\sum_{k=3}^{\infty}
  \left(\frac{24}{k-1}+\frac{36}{k}-\frac{72}{k^{2}}+\frac{72}{k+1}-\frac{12}{k+2}\right)=119-12\pi^{2},
\end{eqnarray}
where we have used the fact that $\sum_{m=1}^{\infty}
\frac{1}{m^{2}}=\frac{1}{6\pi^{2}}$. Substituting Eqs. (\ref{AC1})
and (\ref{AC2}) into Eq. (\ref{ACC2}), we have
\begin{eqnarray}\label{ACC3}
 C = \frac{4}{3}- \left(119-12\pi^{2}\right)\simeq 0.768.
\end{eqnarray}
Thus the average clustering coefficient $C$ of RAN is large and
independent of network size. Since Eqs. (\ref{degree}), (\ref{Ck}),
(\ref{AC1}), and (\ref{AC2}) are exact, it is the same with the
obtained $C$ value. We have performed extensive numerical
simulations of the RAN. In Fig. \ref{clu}, we present the simulation
results about the average clustering coefficient of RAN, which are
in complete agreement with the analytical value.

It should be mentioned that previous jobs~\cite{ZhYaWa05,ZhRoCo05}
have studied the average clustering coefficient of RAN, using
continuum approximation and integral methods, and have acquired a
analytical value of $C=\frac{46}{3}-36\ln\frac{3}{2}\simeq 0.737$.
From Fig. \ref{clu}, we can see that there is a difference between
this analytical value and simulated ones. Where does the difference
come from? And why Eq. (\ref{ACC3}) is more accurate? These might be
explained as follows. In RAN, a node may have only an integer number
of connections (degrees). The continuum approach used
previously~\cite{ZhYaWa05,ZhRoCo05} is under the assumption that the
degree distribution $P(k)$ is modeled by the continuous distribution
$P_{c}(k)=\alpha k^{-3}$ ($k\geq 3$), where $\alpha=18$ is a
normalization constant. It does not properly account for the
fraction (i.e. degree distribution) of nodes with small
degrees~\cite{DoMe01}, and therefore has a discrepancy with the
exact result for RAN with the genuine discrete degree distribution
$P_{d}(k)$ expressed by Eq. (\ref{degree})  [see Fig. \ref{clu}].
However, it is worth noticing that (a) $P_{c}(k)$ and  $P_{d}(k)$
are equal, asymptotically, in the limit of large $k$ and (b) the
difference is most pronounced for $k \approx 3$. The discrepancy is
mainly due to the values of $P(k)$ for small $k$. Thus, the discrete
method, omitting a continuum assumption, seems more suitable for
obtaining exact results for some challenging tasks~\cite{AlBa02}.

%%%%%%%%%%%%%%%%%%%%%%%%%%%%%%%%%%%%%%%%%%%%%%%%%%%%%%%%%%
% Figure  3
%%%%%%%%%%%%%%%%%%%%%%%%%%%%%%%%%%%%%%%%%%%%%%%%%%%%%%%%%%
%%%%%%%%%%%%%%%%%%%%%%%%%%%%%%%%%%%%%%%%%%%%%%%%%%%%%%%%%%

\section{Conclusion}

 %\emph{Conclusion and discussion.}
In this article, we have done an analytical study of the
correlations between the degrees of neighboring nodes in random
Apollonian networks (RAN). Applying rate equation method in the
continuous $k$ approximation, together with its boundary condition
that node $i$ was added to the system at time $i$ with the expected
sum of degrees of its neighbors being $R_i(i)=3+\frac{9}{2}\ln i$,
we have provided the solution of the average degree of nearest
neighbors as the measure of two degree correlations. The obtained
result shows that RAN lacks two degree correlations, which means
that the ANND of nodes with degree $k$ is independent of $k$.
Moreover, we have obtained the exact degree distribution of RAN, on
the basis of which we have attained the exact value of mean
clustering coefficient as an average quantity of three node
correlations. We found that RAN are highly clustered, and the
clustering spectrum scales as $C(k)\sim k^{-1}$, exhibiting
hierarchical organization and modularity. These obtained structure
properties may be helpful to understand and explain the workings of
systems built on RAN. Many physical processes, such as disease
spread and random walks, taking place in RAN exhibit different
behaviors than those in the classic BA networks, which are relevant
to these particular topologies of
RAN~\cite{ZhYaWa05,LiGaHe04,HuXuWuWa06}.

\section*{Acknowledgment}
This research was supported by the National Natural Science
Foundation of China under Grant Nos. 60373019, 60573183, and
90612007, and the Postdoctoral Science Foundation of China under
Grant No. 20060400162. The authors thank Zhen Shen for his
assistance in preparing the manuscript.
%%%%%%%%%%%%%%%%%%%%%%%%%%%%%%%%%%%%%%%%%%%%%%%%%%%%%%%%%%%%%%%%%
%%%%%%%%%%%%%%%%%%%%%%%%%%%%%%%%%%%%%%%%%%%%%%%%%%%%%%%%%%%%%%%%%

\end{document}